\documentclass[sigconf]{acmart}

\copyrightyear{2026}
\acmYear{2026}
\setcopyright{cc}
\setcctype{by}
\acmConference[FSE Companion '26]{34th ACM Joint European Software Engineering Conference and Symposium on the Foundations of Software Engineering}{July 05--09, 2026}{Montreal, QC, Canada}
\acmBooktitle{34th ACM Joint European Software Engineering Conference and Symposium on the Foundations of Software Engineering (FSE Companion '26), July 05--09, 2026, Montreal, QC, Canada}
\acmDOI{10.1145/3803437.3805532}
\acmISBN{979-8-4007-2636-1/2026/07}

\usepackage{enumitem}
\usepackage{hyperref}
\usepackage{tabularx}
\usepackage{xurl}  
\usepackage{graphicx}
\usepackage{amsmath}
\usepackage{multicol}
\usepackage{multirow}
\author{Fariha Tanjim Shifat}
\orcid{0009-0001-9305-5961}
\affiliation{%
  \institution{Missouri University of Science and Technology}
  \city{Rolla, Missouri}
  \country{USA}
}
\email{fsvfh@mst.edu}

\author{Hariswar Baburaj}
\orcid{0009-0000-4663-8112}
\affiliation{%
  \institution{Missouri University of Science and Technology}
  \city{Rolla, Missouri}
  \country{USA}
}
\email{hbkkb@mst.edu}

\author{Ce Zhou}
\orcid{0000-0001-8441-6000}
\affiliation{%
  \institution{Missouri University of Science and Technology}
  \city{Rolla, Missouri}
  \country{USA}
}
\email{cezhou@mst.edu}
 
\author{Jaydeb Sarker}
\orcid{0000-0001-6440-7596}
\affiliation{%
  \institution{University of Nebraska Omaha}
  \city{Omaha, Nebraska}
  \country{USA}
}
\email{jsarker@unomaha.edu}

\author{Mia Mohammad Imran}
\orcid{0000-0003-2477-9971}
\affiliation{%
  \institution{Missouri University of Science and Technology}
  \city{Rolla, Missouri}
  \country{USA}
}
\email{imranm@mst.edu}

\title[LLM-Enabled Open-Source Systems in the Wild: \\An Empirical Study of Vulnerabilities in GitHub Security Advisories]{LLM-Enabled Open-Source Systems in the Wild: An Empirical Study of Vulnerabilities in GitHub Security Advisories}

\begin{abstract}
Large language models (LLMs) are increasingly embedded in open-source software (OSS) ecosystems, creating complex interactions among natural language prompts, probabilistic model outputs, and execution-capable components. However, it remains unclear whether traditional vulnerability disclosure frameworks adequately capture these model-mediated risks. To investigate this, we analyze 295 GitHub Security Advisories published between January 2025 and January 2026 that reference LLM-related components, and we manually annotate a sample of 100 advisories using the OWASP Top 10 for LLM Applications 2025.

We find no evidence of new implementation-level weakness classes specific to LLM systems. Most advisories map to established CWEs, particularly injection and deserialization weaknesses. At the same time, the OWASP-based analysis reveals recurring architectural risk patterns, especially Supply Chain, Excessive Agency, and Prompt Injection, which often co-occur across multiple stages of execution. These results suggest that existing advisory metadata captures code-level defects but underrepresents model-mediated exposure. We conclude that combining the CWE and OWASP perspectives provides a more complete and necessary view of vulnerabilities in LLM-integrated systems.
\end{abstract}


\begin{document}
\maketitle

\sloppy

\section{Introduction}

Large language models (LLMs) are increasingly embedded in modern software systems as development assistants~\cite{hou2024large}, runtime components~\cite{Wang2023ASO}, and autonomous agents~\cite{xi2025rise}. In open-source software (OSS), this integration appears in code generation, dependency management, automated review, conversational interfaces, retrieval-augmented generation (RAG), model context protocol (MCP) pipelines, and tool orchestration frameworks~\cite{he2025llm}.

Unlike conventional software components that operate within fixed input-output structures, LLM-based systems process natural language prompts and produce probabilistic outputs. These outputs can trigger downstream actions such as file system operations, command execution, database queries, or external API calls~\cite{ac1f09077393404a8bea5141d8710259, toolformer}. This coupling between non-deterministic model behavior and execution-capable components introduces additional abstraction layers between user input and system behavior, making system behavior more interaction-driven and harder to reason about statically.

As LLM capabilities are packaged into libraries, inference engines, and agent frameworks distributed through ecosystems such as PyPI and npm, vulnerabilities in these components propagate through the software supply chain~\cite{wang2025large, he2025llm, hu2025understanding}. At the same time, LLM-integrated systems introduce additional exposure pathways through dynamic prompt construction, probabilistic output handling, and autonomous tool invocation~\cite{ferrag2025prompt}. These mechanisms do not map cleanly to conventional package-level risk.

GitHub Security Advisories (GHSAs) document vulnerabilities through structured metadata such as affected packages, version ranges, severity ratings, and Common Weakness Enumeration (CWE) identifiers~\cite{liu2025empirical, github_advisory_database}. CWE classifications capture implementation-level weaknesses such as improper input neutralization, unsafe deserialization, command injection, and uncontrolled resource consumption~\cite{cwe_mitre}. However, they do not indicate whether a weakness is triggered, amplified, or propagated through model reasoning, prompt manipulation, or agent autonomy. This limits their ability to represent how vulnerabilities manifest in LLM-integrated systems.

\begin{figure}[tb]
\centering
\includegraphics[width=\linewidth]{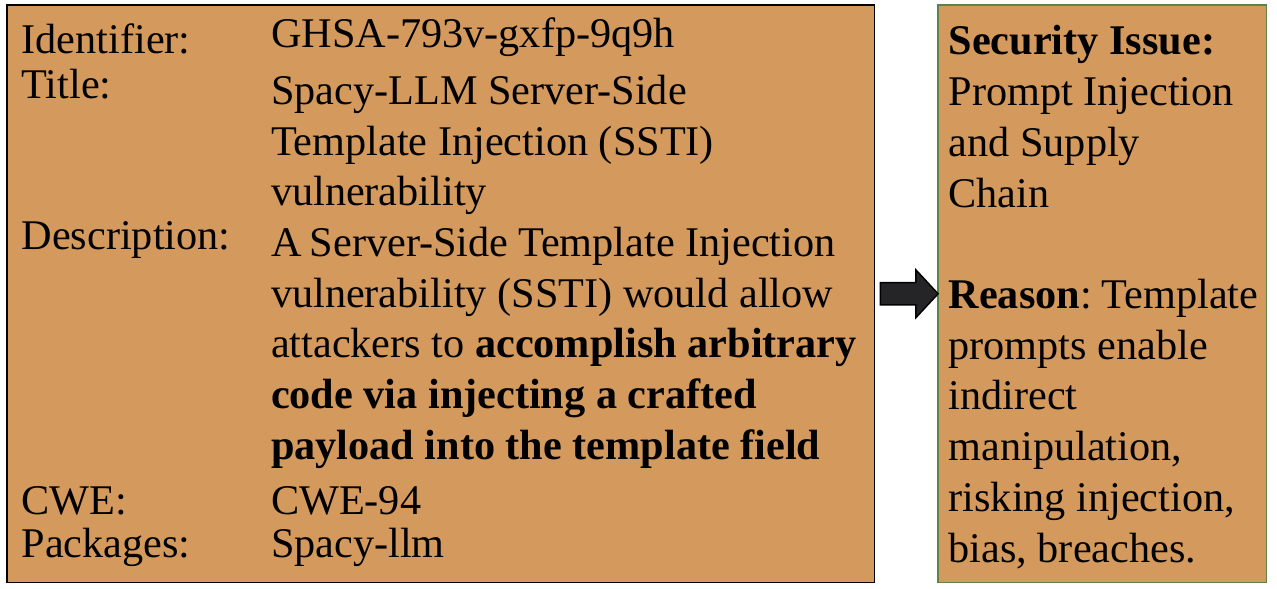}
\caption{Illustration of model-mediated exposure in an LLM-associated advisory.}
\label{fig:motivation}
\end{figure}

Figure~\ref{fig:motivation} illustrates this limitation. The advisory is correctly classified as CWE-94 (improper control of code generation), which captures the code-level defect. However, the exploitation pathway operates through template-based prompts and model-generated content. These indirect manipulation channels are not represented by the CWE label. The vulnerability, therefore, exists at two levels: the implementation defect captured by CWE and the interaction pattern through which model outputs reach executable behavior. The GHSA schema captures only the implementation level defects, leaving the architectural risk entirely unrepresented.

The OWASP Top 10 for LLM Applications 2025 (Table~\ref{tab:categories}) provides a complementary perspective~\cite{owasp-llm-top10-2025}. It defines categories such as Prompt Injection, Improper Output Handling, Excessive Agency, Data and Model Poisoning, and Unbounded Consumption. These categories capture interaction-level risks. CWE explains how code fails, whereas OWASP captures how model-enabled systems become exposed at the architectural level.

Despite the growing prevalence of LLM-integrated software, empirical evidence on how LLM-related vulnerabilities appear in open-source advisories remains limited. It remains unclear how often LLM-associated packages appear in disclosures, whether CWE classifications adequately reflect model-mediated exposure, and how advisory data aligns with LLM-specific risk taxonomies. This analysis is further complicated by the fact that current advisory schemas do not include structured indicators of LLM involvement.

To address this gap, we conduct an empirical study of 295 GitHub Security Advisories published between January 2025 and January 2026 that explicitly reference LLM-related components. From this dataset, we manually categorize the 133 unique affected packages into three groups: \textit{LLM-associated}, \textit{Possible LLM-associated}, and \textit{Non-LLM-associated}. The first group includes packages that directly implement or orchestrate LLM functionality, such as inference, prompt management, and agent frameworks. The second includes packages commonly used within LLM pipelines without being LLM-specific by design. The third includes packages whose functionality is unrelated to LLM systems. We then randomly sample 100 advisories from the first two categories and manually annotate them using the OWASP Top 10 for LLM Applications 2025~\cite{owasp-llm-top10-2025} to examine architectural exposure patterns beyond implementation-level defects. Specifically, we ask the following RQs:

\noindent
\textbf{RQ1}: Which CWE categories most commonly occur in GitHub Security Advisories related to LLM-associated packages?

$\to$ Code injection (\textit{CWE-94}), command injection (\textit{CWE-77}, \textit{CWE-78}), and unsafe deserialization (\textit{CWE-502}) dominate, indicating that LLM ecosystems primarily inherit established weakness classes.

\noindent
\textbf{RQ2}: How effectively do existing advisory metadata fields represent LLM involvement and model-mediated exposure mechanisms?

$\to$ Current GHSA metadata lacks structured indicators of LLM involvement, requiring manual classification to identify model-mediated exposure patterns.

\noindent
\textbf{RQ3}: What exposure patterns emerge when LLM-related advisories are mapped using the OWASP Top 10 for LLM Applications 2025?

$\to$ Supply Chain Risks (44\%), Excessive Agency (20\%), and Prompt Injection (18\%) dominate. Thirty-seven percent of advisories exhibit multi-label patterns that combine prompt manipulation, output handling weaknesses, and execution authority.

\noindent
\textbf{RQ4}: How do OWASP LLM risk categories correspond to underlying CWE weakness classes?

$\to$ Architectural LLM risk categories consistently materialize through conventional implementation weaknesses. Supply Chain shows wide CWE diversity, whereas Excessive Agency and Prompt Injection concentrate on injection-related flaws.

%Together, these findings make three contributions:
We have following three contributions from this study:

\begin{enumerate}
\item an empirical characterization of CWE patterns in LLM-associated GitHub Security Advisories, based on 295 disclosures from January 2025 to January 2026;
\item a systematic mapping between CWE and the OWASP Top 10 for LLM Applications 2025 through manual annotation of 100 advisories; and,
\item evidence that current advisory metadata lacks explicit indicators of LLM involvement, which limits systematic analysis of model-mediated exposure.
\end{enumerate}

The replication package, including annotation guidelines and the annotated dataset, is publicly available at~\cite{replication-LLMSC-2026}. The remainder of the paper is organized as follows. Section~\ref{sec:background} reviews related work. Section~\ref{sec:method} describes the methodology. Section~\ref{sec:evaluation} presents the results. Section~\ref{sec:discussion} discusses the implications. Section~\ref{sec:limitation} outlines the limitations. Section~\ref{sec:conclusion} concludes the paper.

\begin{figure*}[tb]
\centering
\includegraphics[width=\linewidth]{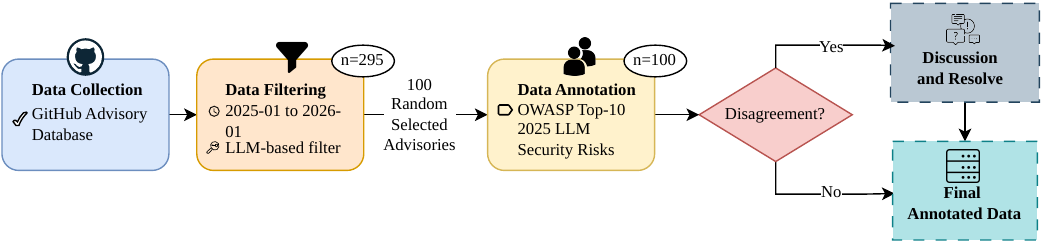}
\caption{Overview of the data collection, filtering, sampling, and annotation pipeline.}
\label{fig:method}
\end{figure*}

\begin{table*}[tb]
\centering
\caption{OWASP TOP 10 LLM Risk Categories in 2025.}
\label{tab:categories}

\begin{tabular}{|l|p{4.7cm}|p{10.3cm}|}
\hline
\textbf{ID} & \textbf{Category Name} & \textbf{Definition} \\
\hline
LLM01 & Prompt Injection & Inputs manipulate LLM behavior or outputs beyond intended controls. \\
\hline
LLM02 & Sensitive Information Disclosure & Unintended exposure of confidential, personal, or proprietary data. \\
\hline
LLM03 & Supply Chain & Risks from compromised models, data, tools, or dependencies in the LLM lifecycle. \\
\hline
LLM04 & Data and Model Poisoning & Malicious manipulation of training or embedding data to alter model behavior. \\
\hline
LLM05 & Improper Output Handling & Unsafe use of LLM outputs without validation or sanitization. \\
\hline
LLM06 & Excessive Agency & LLMs granted excessive autonomy or permissions causing harmful actions. \\
\hline
LLM07 & System Prompt Leakage & Exposure of system prompts containing sensitive instructions or data. \\
\hline
LLM08 & Vector and Embedding Weaknesses & Flaws in vector or embedding handling enabling data leakage or manipulation. \\
\hline
LLM09 & Misinformation & Generation of incorrect or misleading information that appears credible. \\
\hline
LLM10 & Unbounded Consumption & Uncontrolled LLM usage leading to resource exhaustion or financial loss. \\
\hline
\end{tabular}

\end{table*}

\section{Background and Related Work}
\label{sec:background}

Our study draws on two related areas: open source software supply chain security and security risks in LLM-enabled systems.

\subsection{Open Source Supply Chain Security}

Modern software development relies heavily on third-party open source packages distributed through registries such as npm, PyPI, and Maven. When developers install a package, they also introduce its transitive dependencies, which expands the attack surface. As a result, a single malicious or vulnerable package can propagate risk across thousands of downstream projects~\cite{zimmermann2019small, ohm2020backstabber}.

Prior work has examined the scale and dynamics of these risks across ecosystems. Alfadel \textit{et} al.~\cite{alfadel2023empirical} analyzed 1,396 vulnerability reports affecting 698 Python packages and found that vulnerabilities often take more than three years to surface, while more than half remain unfixed at the time of public disclosure. Zimmermann \textit{et} al.~\cite{zimmermann2019small} showed that the npm ecosystem has a ``small world with high risks'' topology, in which a small number of highly connected maintainer accounts can affect large portions of the registry. Decan \textit{et} al.~\cite{decan2018impact} examined how vulnerabilities propagate through npm dependency networks and highlighted the cascading nature of supply chain exposure. Guo \textit{et} al.~\cite{guo2023empirical} studied malicious code in the PyPI ecosystem and identified thousands of malicious packages that relied on typosquatting and related social engineering techniques.

To support the management of these threats, the GHSA database provides structured vulnerability metadata, including affected packages, version ranges, and CWE identifiers~\cite{GitHubAdvisoryDatabase}. However, Segal \textit{et} al.~\cite{segal2025characterizing} showed that structural latencies in the GHSA review pipeline can delay the availability of actionable security information. This means that metadata alone may leave temporal windows of exposure. Complementary code-centric approaches, such as the detection and mitigation frameworks proposed by Ponta \textit{et} al.~\cite{ponta2020detection}, therefore remain important for securing open source dependencies.

\subsection{LLM Supply Chain and Security Risks}

The rapid adoption of LLMs has expanded the notion of software supply chains beyond conventional package dependencies to include pre-trained models, training datasets, fine-tuning pipelines, prompt templates, and orchestration frameworks~\cite{wang2025large}. In response, recent work has proposed broader research agendas~\cite{wang2025large} and taxonomies of LLM-specific threats, including malicious package injection, model backdoors, and prompt leakage~\cite{huang2024lifting}.

Researchers have also shown that LLM interfaces can be actively exploited through prompt injection. These attacks have been demonstrated empirically and formalized in attack frameworks~\cite{greshake2023not, liu2023prompt, liu2024formalizing}. More recent work extends these concerns to autonomous agent ecosystems and highlights severe protocol-layer vulnerabilities~\cite{ferrag2025prompt}.

Despite this growing body of research, little work has systematically mapped GitHub Security Advisories involving LLM-associated packages to LLM security risks. Our study addresses this gap by connecting implementation-level weakness reporting with architectural exposure analysis.

\section{Methodology}
\label{sec:method}

Figure~\ref{fig:method} presents the overall methodology of this study. In this section, we describe the procedures used to collect advisory data, filter for LLM-associated packages, and conduct manual annotation.

\subsection{Data Collection and Filtering}
We collected publicly available security advisories from the GitHub Advisory Database (GHSA) published between January 2025 and January 2026~\cite{GitHubAdvisoryDatabase}. This period provides a recent and consistent snapshot of disclosures during rapid LLM ecosystem integration. GHSA standardizes records across major open-source ecosystems on GitHub, including projects that distribute LLM frameworks, inference engines, and orchestration libraries. It therefore offers direct visibility into supply chain risks affecting AI-enabled software on GitHub. Prior work shows that GHSA is effective for characterizing supply chain vulnerabilities~\cite{segal2025characterizing}, particularly those driven by dependency structures in open-source software stacks~\cite{liu2025empirical}.

GHSA metadata includes affected packages, ecosystems, severity ratings, version ranges, CVE identifiers, CWE classifications, and vulnerability descriptions. For each advisory, we extracted the GHSA identifier, description, affected ecosystem and package names, severity level, CVE identifier when available, and associated CWE categories.

To identify LLM-relevant cases, we applied keyword-based filtering to advisory metadata and descriptions. The keyword set included generic LLM terms (e.g., llm, gpt, embeddings), model vendors and families (e.g., openai, llama, mistral, anthropic, claude), orchestration and agent frameworks (e.g., langchain, flowise, ollama, vllm, huggingface), protocol-related terms (e.g., mcp, model context protocol), and retrieval or embedding infrastructure (e.g., vector databases, chroma, milvus). This process yielded 295 advisories referencing LLM-related components.

\subsection{LLM-associated Package Filtering}
Keyword matches alone do not guarantee that an affected package directly implements LLM functionality. In several cases, advisories referenced LLM concepts in their descriptions while affecting general-purpose libraries or supporting infrastructure. To refine the dataset, we extracted all affected package names from the 295 disclosures, resulting in 133 unique packages. We then manually reviewed each package using its repository documentation, project description, and stated functionality.

We grouped the packages into three categories using thematic analysis~\cite{clarke2017thematic}: \textit{LLM-associated}, \textit{Possible LLM-associated}, and \textit{Non-LLM-associated}. One author performed the primary classification based on package names, descriptions, and documentation. A second author independently reviewed the assignments. Disagreements were resolved through discussion. Section~\ref{rq2} provides the detailed definitions of each category.

The final distribution included 84 \textit{LLM-associated} packages affecting 226 advisories, 16 \textit{Possible LLM-associated} packages affecting 34 advisories, and 33 \textit{Non-LLM-associated} packages affecting 35 advisories. We then randomly sampled 100 advisories from the 260 associated with the first two categories, yielding 92 \textit{LLM-associated} and 8 \textit{Possible LLM-associated} advisories for detailed examination.

\subsection{Data Annotation}
Two authors independently annotated the randomly selected 100 advisories. We developed a detailed annotation guideline based on the OWASP Top 10 for LLM Applications 2025 taxonomy~\cite{owasp-llm-top10-2025}. This taxonomy includes Prompt Injection, Sensitive Information Disclosure, Supply Chain, Data and Model Poisoning, Improper Output Handling, Excessive Agency, System Prompt Leakage, Vector and Embedding Weaknesses, Misinformation, and Unbounded Consumption. Table~\ref{tab:categories} summarizes these definitions. The full annotation instructions are available in the replication package~\cite{replication-LLMSC-2026}.

Two annotators independently reviewed all advisories. For each case, they analyzed the advisory content, examined the technical role of the affected package, and assessed the exploitation scenario. They then identified the dominant exposure mechanism and mapped the advisory to one or more OWASP categories. When an advisory did not correspond to any of the ten predefined categories, it was labeled as miscellaneous.

We measured inter-annotator agreement before resolving discrepancies. Cohen’s Kappa~\cite{cohen1960} was 0.76 and Gwet’s AC1~\cite{gwet2014handbook} was 0.95, indicating substantial agreement and high reliability. The annotators then reconciled differences through discussion and finalized the labels used in the subsequent analyses.

\section{Evaluation and Results}
\label{sec:evaluation}

To systematically characterize LLM-related vulnerabilities within the GitHub Security Advisory ecosystem, we organize our empirical findings around our four research questions.

\begin{table*}[tb]
\centering
\caption{Top 10 CWEs in LLM-Referenced Advisories.}
\label{tab:cwe}
\begin{tabular}{|l|p{10cm}|c|l|}
\hline
\textbf{CWE} & \textbf{Description} & \textbf{Count} & \textbf{Example} \\
\hline
CWE-94 & Improper Control of Generation of Code (Code Injection) & 24 & \texttt{GHSA-793v-gxfp-9q9h}~\cite{ghsa-793v-gxfp-9q9h} \\
CWE-502 & Deserialization of Untrusted Data & 22 & \texttt{GHSA-mrw7-hf4f-83pf}~\cite{ghsa-mrw7-hf4f-83pf} \\
CWE-77 & Improper Neutralization of Special Elements used in a Command (Command Injection) & 22 & \texttt{GHSA-xq4m-mc3c-vvg3}~\cite{ghsa-xq4m-mc3c-vvg3} \\
CWE-78 & Improper Neutralization of Special Elements used in an OS Command (OS Command Injection) & 19 & \texttt{GHSA-2vv2-3x8x-4gv7}~\cite{ghsa-2vv2-3x8x-4gv7} \\
CWE-79 & Improper Neutralization of Input During Web Page Generation (Cross-site Scripting) & 19 & \texttt{GHSA-hfcf-79gh-f3jc}~\cite{ghsa-hfcf-79gh-f3jc} \\
CWE-22 & Improper Limitation of a Pathname to a Restricted Directory (Path Traversal) & 18 & \texttt{GHSA-j9g7-mqhh-9hxf}~\cite{ghsa-j9g7-mqhh-9hxf} \\
CWE-1333 & Inefficient Regular Expression Complexity & 14 & \texttt{GHSA-8r9q-7v3j-jr4g}~\cite{ghsa-8r9q-7v3j-jr4g} \\
CWE-770 & Allocation of Resources Without Limits or Throttling & 12 & \texttt{GHSA-hf3c-wxg2-49q9}~\cite{ghsa-hf3c-wxg2-49q9} \\
CWE-400 & Uncontrolled Resource Consumption & 11 & \texttt{GHSA-6fvq-23cw-5628}~\cite{ghsa-6fvq-23cw-5628} \\
CWE-89 & Improper Neutralization of Special Elements used in an SQL Command (SQL Injection) & 11 & \texttt{GHSA-jmgm-gx32-vp4w}~\cite{GHSA-jmgm-gx32-vp4w} \\
\hline
\end{tabular}

\end{table*}

\subsection{RQ1: Which CWE categories most commonly occur in GitHub Security Advisories related to LLM-focused packages?}

We address RQ1 by analyzing CWE distributions in both the full 295-advisory dataset and the 100-advisory annotated subset. Across the full dataset, we observe 99 distinct CWE identifiers. Table~\ref{tab:cwe} reports the ten most frequent CWE categories. \textit{CWE-94} appears most often, with 24 occurrences, followed by \textit{CWE-77} and \textit{CWE-502}, with 22 occurrences each.

These CWEs fall into three broad weakness types. First, injection-related flaws form the largest group. This group includes code injection (\textit{CWE-94}), command injection (\textit{CWE-77}, \textit{CWE-78}), SQL injection (\textit{CWE-89}), and cross-site scripting (\textit{CWE-79}). Second, unsafe handling of untrusted data appears through deserialization (\textit{CWE-502}) and path traversal (\textit{CWE-22}). Third, resource and complexity management weaknesses, including inefficient regular expression evaluation (\textit{CWE-1333}), uncontrolled resource allocation (\textit{CWE-770}), and uncontrolled resource consumption (\textit{CWE-400}), expose systems to denial-of-service conditions.

Several advisories illustrate how these weaknesses appear in LLM-enabled systems. \texttt{GHSA-793v-gxfp-9q9h} (Figure~\ref{fig:motivation}) describes a server-side template injection in spacy-llm that enables arbitrary code execution, corresponding to \textit{CWE-94}~\cite{ghsa-793v-gxfp-9q9h}. \texttt{GHSA-mrw7-hf4f-83pf} reports unsafe deserialization in vLLM~\cite{ghsa-mrw7-hf4f-83pf}, consistent with \textit{CWE-502}.

The 100-advisory subset shows the same overall pattern. Injection-related weaknesses remain dominant, with \textit{CWE-77} appearing 11 times, \textit{CWE-78} and \textit{CWE-502} appearing 9 times each, and \textit{CWE-94} appearing 8 times. Authentication and access control weaknesses, particularly \textit{CWE-306}, appear more prominently in this subset. No distinct LLM-specific implementation weakness emerges.

Overall, the vulnerability patterns align with conventional software security trends. The data suggests continuity rather than divergence, with established weakness classes persisting in LLM-enabled architectures.

\subsection{RQ2: How effectively do existing advisory metadata fields represent LLM involvement and model-mediated exposure mechanisms?}\label{rq2}

We investigate RQ2 through two lenses: the structured metadata in the 295 filtered advisories and the results of manual package categorization. Table~\ref{tab:issue_dist} summarizes ecosystem and severity distributions, and Table~\ref{tab:packages} presents package-level information.

The advisory schema includes ecosystem registry, severity level, fix status, and affected package name. PyPI contributes 162 advisories, npm 96, Go 22, Packagist 10, and crates.io and Maven 3 each. For severity, PyPI includes 31 Critical, 69 High, 53 Moderate, and 9 Low cases. npm includes 24 Critical, 50 High, 18 Moderate, and 4 Low. Go includes 3 Critical, 11 High, 6 Moderate, and 2 Low. Packagist includes 1 Critical, 2 High, and 7 Moderate. crates.io includes 3 Low. Maven includes 1 High and 2 Moderate.

These fields support cross-ecosystem and severity-level comparisons. However, they do not indicate whether a vulnerability arises in an LLM-enabled component. The metadata does not distinguish conventional software libraries from packages that implement model inference, prompt handling, or agent pipelines.

% \todo{@SHIFAT: Add the definitions of these 3 categories and their boundary conditions.}

To evaluate the relevance of LLMs, we conducted a thorough review of all 133 unique packages in the dataset. This involved analyzing repository documentation, project descriptions, API references, and example usage to determine whether LLM functionality was a core feature, a supporting integration, or unrelated to LLM systems. Based on our analysis, we classified the packages into three categories:

\begin{itemize}
    \item \textit{LLM-associated}: Packages whose primary purpose is to implement 
    or orchestrate LLM functionality, including model inference engines, prompt 
    management libraries, agent frameworks, and retrieval-augmented generation 
    pipelines. If removing LLM support eliminates the package's core purpose or 
    renders it non-functional, we classify it as LLM-associated.
    
    \item \textit{Possible LLM-associated}: Packages that do not inherently 
    implement LLM functionality but serve as common supporting infrastructure 
    in LLM-based systems, such as vector databases, workflow orchestration tools, 
    and data processing utilities. If the package remains fully functional outside 
    LLM contexts but frequently appears in LLM pipelines, we classify it as 
    Possible LLM-associated.
    
    \item \textit{Non LLM-associated}: Packages whose core functionality is 
    unrelated to LLM systems. If the package's main purpose is general software 
    functionality and its connection to LLM-enabled systems is weak, indirect, 
    or non-essential, we classify it as Non-LLM-associated.
\end{itemize}

Table~\ref{tab:packages} shows the distribution of advisories in each package categories. 84 packages directly implement or orchestrate LLM functionality and account for 226 advisories. Sixteen packages are \textit{possible LLM-associated} and account for 34 advisories. Thirty-three packages are \textit{non-LLM-associated} and account for 35 advisories.

Overall, the existing advisory metadata captures ecosystem distribution and severity levels, but it does not structurally encode generative AI involvement or model-mediated exposure mechanisms.

\begin{table}[tb]
\centering
\caption{Distribution of security issues per ecosystem by severity and fix status (at the time of data collection).}
\label{tab:issue_dist}

\begin{tabular}{|l|cccc|cc|}
\hline
\multirow{2}{*}{\textbf{Ecosystem}} & \multicolumn{4}{c|}{\textbf{Severity}} & \multicolumn{2}{c|}{\textbf{Resolved}} \\ 
 & Critical & High & Moderate & Low & Yes &No \\
\hline
PyPI & 31 & 69 & 53 & 9 & 132 & 30\\
npm & 24 & 50 & 18 & 4 & 78 & 18\\
Go & 3 & 11 & 6 & 2 & 12 & 10\\
Packagist & 1 & 2 & 7 & 0 & 9 & 1\\
crates.io & 0 & 0 & 0 & 3 & 3 & 0\\
Maven & 0 & 1 & 2 & 0 & 3 & 0\\
\hline
% Total & 59 & 133 & 86 & 18 & 255 & 59 \\
% \hline
\end{tabular}

\end{table}

\begin{table}[tb]
\centering

\caption{Package categorization and advisory distribution across 295 advisories}
\label{tab:packages}
\begin{tabular}{|p{2cm}|c|c|p{2cm}|}
\hline
\textbf{Category} & \textbf{Packages} & \textbf{Advisories} & \textbf{Examples} \\
\hline
\multirow{3}{*}{LLM-associated} & \multirow{3}{*}{84} & \multirow{3}{*}{226} & vllm, ollama, langchain, litellm \\
\hline
Possible LLM-associated & \multirow{2}{*}{16} & \multirow{2}{*}{34} & keras, milvus, weaviate, n8n \\
\hline
Non-LLM-associated &  \multirow{2}{*}{33} & \multirow{2}{*}{35} & wasmtime, orejime, directus \\
\hline
\end{tabular}

\end{table}

% \begin{table*}[tb]
% \centering

% \caption{Package categorization and advisory distribution across 295 advisories}
% \label{tab:packages}
% \begin{tabular}{|p{2cm}|p{8.3cm}|c|c|p{2cm}|}
% \hline
% \textbf{Category} & \textbf{Definition} & \textbf{Packages} & \textbf{Advisories} & \textbf{Examples} \\
% \hline
% LLM-associated & Packages that directly implement, integrate, or orchestrate large language models, including model inference, prompt management, agent frameworks, MCP infrastructure, and LLM-centric APIs. & 84 & 226 & vllm, ollama, langchain, litellm \\
% \hline
% Possible LLM-associated & Packages that are not LLM-specific by design but are commonly used within LLM pipelines or indirectly support LLM development, deployment, or integration. & 16 & 34 & keras, milvus, weaviate, n8n \\
% \hline
% Non-LLM-associated & Packages whose dominant functionality is general-purpose infrastructure, system software, security tooling, or unrelated utilities rather than LLM-specific implementation or orchestration. & 33 & 35 & wasmtime, orejime, directus \\
% \hline
% \end{tabular}

% \end{table*}

\begin{figure}[tb]
\centering
\includegraphics[width=0.95\linewidth]{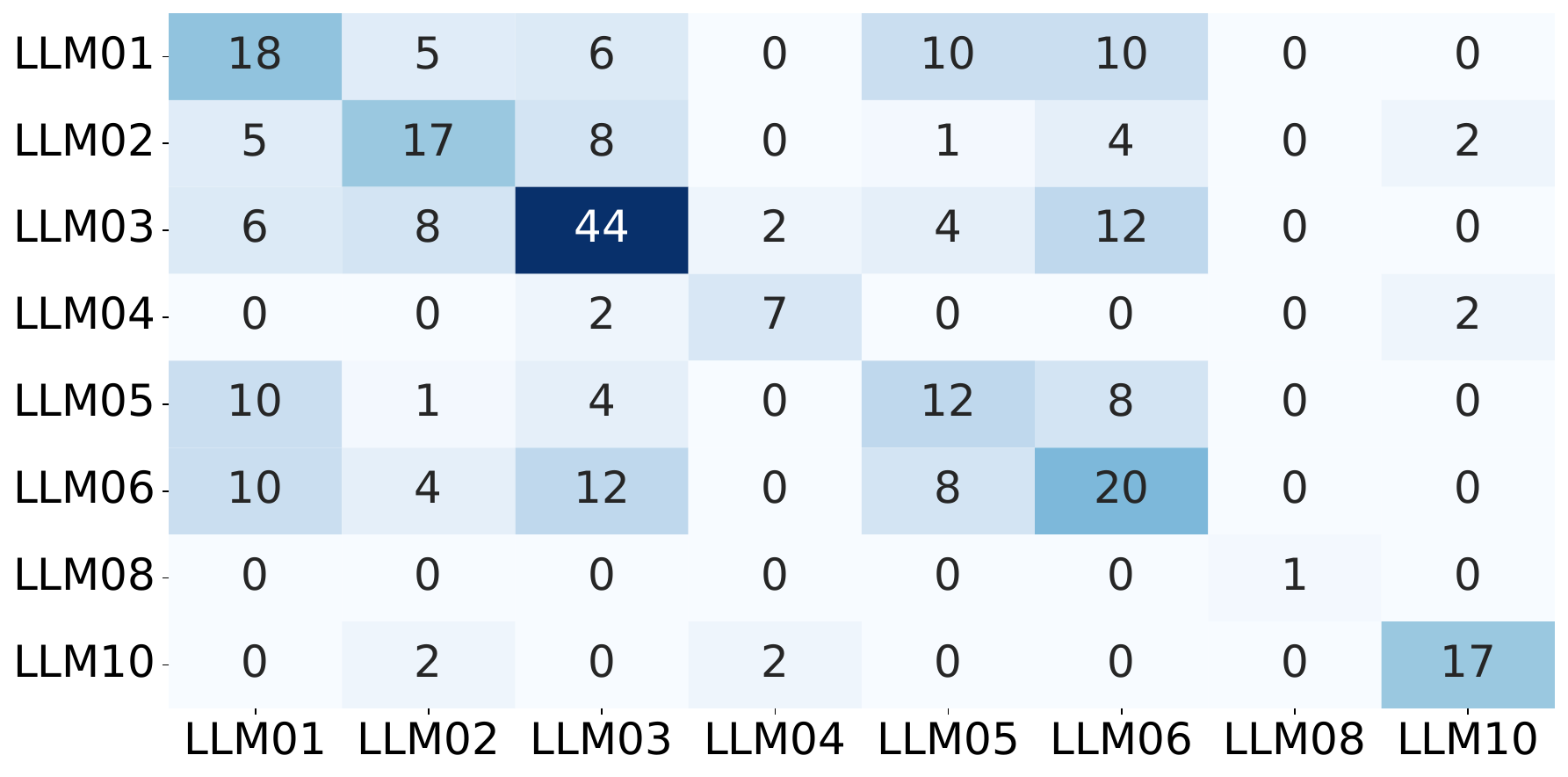}
\caption{Co-occurrence of OWASP LLM Risk Categories.}
\label{fig:heatmap}
\end{figure}

\subsection{RQ3: What exposure patterns emerge when LLM-related advisories are mapped using the OWASP Top 10 for LLM Applications 2025?}

Because advisory metadata does not indicate whether a vulnerability involves LLM functionality, we applied the OWASP Top 10 for LLM Applications 2025 taxonomy~\cite{owasp-llm-top10-2025} to 100 randomly sampled and manually annotated advisories. As described in Section~\ref{sec:method}, we mapped each advisory to one or more OWASP categories. This analysis reveals exposure patterns that are not captured by the standard GHSA advisory schema.

% \input{tables/llm_count}
% Table~\ref{tab:llm_count} shows the resulting distribution. 

\textit{LLM03} (Supply Chain) is the most frequent category, appearing in 44 advisories. This suggests that exposure is concentrated in dependencies, plugins, orchestration layers, and surrounding infrastructure rather than in model internals alone. For example, \texttt{GHSA-793v-gxfp-9q9h} shows how a weakness in an ecosystem component becomes exploitable within an LLM-driven workflow~\cite{ghsa-793v-gxfp-9q9h}.

\textit{LLM06} (Excessive Agency) appears in 20 advisories and captures architectures in which LLM-mediated components can trigger execution paths or invoke privileged actions. In \texttt{GHSA-3ch2-jxxc-v4xf}, model-influenced tool input reaches an execution interface and enables over-privileged integration~\cite{ghsa-3ch2-jxxc-v4xf}. \textit{LLM01} (Prompt Injection) appears in 18 advisories and often serves as the initiating condition for downstream impact. \textit{LLM02} (Sensitive Information Disclosure) and \textit{LLM10} (Unbounded Consumption) each appear in 17 advisories, reflecting confidentiality exposure and resource abuse, respectively. \textit{LLM05} (Improper Output Handling ) appears in 12 advisories and captures unsafe consumption of model outputs. \textit{LLM04} (Data and Model Poisoning) appears in 7 advisories and primarily affects retrieval-related components. \textit{LLM08} (Vector and Embedding Weaknesses) appears once. No advisories in our sample map to \textit{LLM07} (System Prompt Leakage) or \textit{LLM09} (Misinformation), which may reflect reporting limitations in sampled dataset rather than the absence of these risks.

As shown in Figure~\ref{fig:heatmap}, these risks frequently appear as multi-step combinations rather than isolated categories. Sixty-three advisories carry a single label, whereas 37 carry multiple labels. The pairing \textit{LLM03+LLM06} appears 12 times, indicating that integration weaknesses become more severe when systems expose execution authority. For example, \texttt{GHSA-7944-7c6r-55vv} shows how compromise in an orchestration component can propagate into a server-side execution path once the system provides privileged interfaces~\cite{ghsa-7944-7c6r-55vv}.

Figure~\ref{fig:heatmap} also highlights recurring prompt-driven combinations. \textit{LLM01+LLM05} and \textit{LLM01+LLM06} each appear 10 times, and \textit{LLM01+LLM05+LLM06} appears 7 times. \texttt{GHSA-3ch2-jxxc-v4xf} illustrates this sequence: attacker-controlled prompt input influences tool arguments, weak output handling fails to constrain the response, and the system forwards those parameters to an execution endpoint~\cite{ghsa-3ch2-jxxc-v4xf}. Similarly, \texttt{GHSA-6f6r-m9pv-67jw} reflects the same prompt-to-tool-to-command pathway in a separate MCP deployment~\cite{ghsa-6f6r-m9pv-67jw}.

Overall, these results show that LLM-related exposure often follows recurring architectural chains rather than isolated failures. Prompt processing, output handling, and execution authority interact in ways that create predictable exploitation pathways. In our dataset, the dominant risks arise from cross-component interaction more often than from model internals alone.

\subsection{RQ4: How do OWASP LLM risk categories correspond to underlying CWE weakness classes?}

RQ3 identifies the dominant exposure patterns at the OWASP category level. RQ4 examines how these architectural risk categories correspond to implementation-level CWE weakness classes. We therefore analyze the mapping between OWASP LLM categories and CWE identifiers within the same 100 annotated advisories.

The annotated dataset contains 55 unique CWE identifiers. Eight of the ten OWASP LLM categories appear in the sample. \textit{LLM07} and \textit{LLM09} do not appear, consistent with the distribution reported in RQ3. Table~\ref{tab:llm_cwe_pairs} presents the OWASP-CWE mappings and their frequencies.

\begin{table*}[tb]
\centering
\caption{LLM-CWE Pair Frequencies in the Annotated Sample (Count $\geq$ 4).}
\label{tab:llm_cwe_pairs}
\begin{tabular}{|p{6cm}|p{7cm}|c|}
\hline
\textbf{LLM} & \textbf{CWE} & \textbf{Count} \\
\hline
LLM03 (Supply Chain) & CWE-78 (OS Command Injection) & 9 \\
LLM06 (Excessive Agency) & CWE-77 (Command Injection) & 8 \\
LLM03 (Supply Chain) & CWE-306 (Missing Authentication for Critical Function) & 6 \\
LLM03 (Supply Chain) & CWE-94 (Code Injection) & 5 \\
LLM03 (Supply Chain) & CWE-77 (Command Injection) & 5 \\
LLM03 (Supply Chain) & CWE-502 (Deserialization of Untrusted Data) & 5 \\
LLM06 (Excessive Agency) & CWE-78 (OS Command Injection) & 5 \\
LLM01 (Prompt Injection) & CWE-77 (Command Injection) & 5 \\
LLM10 (Unbounded Consumption) & CWE-770 (Allocation of Resources Without Limits) & 5 \\
LLM01 (Prompt Injection) & CWE-79 (Cross-Site Scripting) & 4 \\
LLM05 (Improper Output Handling) & CWE-77 (Command Injection) & 4 \\
LLM04 (Data and Model Poisoning) & CWE-502 (Deserialization of Untrusted Data) & 4 \\
LLM03 (Supply Chain) & CWE-89 (SQL Injection) & 4 \\
% LLM01 (Prompt Injection) & CWE-78 (OS Command Injection) & 3 \\
% LLM01 (Prompt Injection) & CWE-94 (Code Injection) & 3 \\
% LLM05 (Improper Output Handling) & CWE-78 (OS Command Injection) & 3 \\
% LLM05 (Improper Output Handling) & CWE-94 (Code Injection) & 3 \\
% LLM06 (Excessive Agency) & CWE-94 (Code Injection) & 3 \\
% LLM10 (Unbounded Consumption) & CWE-1333 (Inefficient Regular Expression Complexity) & 2 \\
% LLM10 (Unbounded Consumption) & CWE-400 (Uncontrolled Resource Consumption) & 2 \\
% LLM06 (Excessive Agency) & CWE-306 (Missing Authentication for Critical Function) & 2 \\
% LLM06 (Excessive Agency) & CWE-862 (Missing Authorization) & 2 \\
% LLM02 (Sensitive Information Disclosure) & CWE-79 (Cross-Site Scripting) & 2 \\
\hline
\end{tabular}
\end{table*}

\textit{LLM03} spans 37 distinct OWASP-CWE co-occurrence pairs across 44 advisories, making it the broadest category. The most frequent mapping, \textit{LLM03-CWE-78} (9), indicates OS command injection introduced through third-party LLM integrations. \textit{LLM03-CWE-306} (6) reflects missing authentication in exposed LLM-connected services. Additional recurring mappings, including \textit{LLM03-CWE-94}, \textit{LLM03-CWE-77}, and \textit{LLM03-CWE-502} (5 each), show that supply-chain risks commonly manifest through injection and deserialization weaknesses.

\textit{LLM06} primarily corresponds to execution-related CWEs. \textit{LLM06-CWE-77} (8) and \textit{LLM06-CWE-78} (5) indicate command injection enabled by over-privileged agent behavior. \textit{LLM06-CWE-94} (3) reflects unsafe dynamic code execution. \textit{LLM01} maps mainly to injection-related weaknesses, including \textit{LLM01-CWE-77} (5) and \textit{LLM01-CWE-79} (4), showing how prompt manipulation propagates into command or script injection contexts.

\textit{LLM05} also aligns with execution-related weaknesses, particularly \textit{LLM05-CWE-77} (4) and \textit{LLM05-CWE-78} (3), indicating unsafe consumption of model outputs. \textit{LLM10} corresponds primarily to resource exhaustion, dominated by \textit{LLM10-CWE-770} (5). \textit{LLM04} most frequently maps to \textit{LLM04-CWE-502} (4), reflecting deserialization risks.

Overall, the analysis shows that LLM integration changes how traditional weaknesses interact and propagate across system components. The underlying implementation flaws remain conventional, but architectural coupling shapes their exposure and impact.
\section{Implications}
\label{sec:discussion}

Our findings suggest three main implications for understanding vulnerabilities in LLM-associated software.

\textbf{Implementation-Level Weaknesses Remain Central.}
The results suggest that vulnerabilities in LLM-integrated open-source systems should be analyzed at multiple levels. At the implementation level, the dominant defects remain familiar. The most common advisories map to established CWE families such as injection, deserialization, authentication, and resource management. This continuity indicates that existing code-level security analysis and mitigation techniques remain relevant in LLM-integrated systems. However, CWE labels alone do not explain how these weaknesses become exposed within model-mediated workflows or how they participate in broader interaction sequences.

\textbf{Architectural Exposure Shapes Risk.}
The OWASP-based annotation further indicates that risk in LLM-integrated systems is often shaped by system architecture rather than by model internals alone. The most frequent patterns involve Supply Chain, Excessive Agency, and Prompt Injection, and many advisories exhibit combinations of these categories. This suggests that vulnerabilities often propagate across prompt processing, output handling, orchestration, and execution layers. In this setting, the security question is not only which weakness exists, but also how system design allows that weakness to travel across components and produce downstream effects.

\textbf{Advisory Metadata Captures Only Part of the Exposure Path.}
The results also indicate a limitation in current advisory reporting practices. GHSA metadata records affected packages, severity, and implementation-level weakness classes, but it does not explicitly encode whether a vulnerability involves LLM functionality or model-mediated interaction. As a result, advisory records capture the code-level defect while often omitting the architectural pathway through which the defect becomes exploitable in an LLM-integrated system. This gap suggests that CWE and OWASP provide complementary views of risk: one captures the underlying weakness, while the other captures the interaction pattern through which that weakness is exposed.

\section{Limitations}
\label{sec:limitation}

Our study has several limitations. We restrict the dataset to GitHub Security Advisories published within a defined time window, excluding vulnerabilities disclosed through other channels such as vendor advisories or independent CVE entries. We identify LLM-associated cases through keyword-based filtering, which may omit relevant advisories that do not explicitly reference LLM terminology and may include indirectly related cases. We also rely on manual annotation for OWASP category mapping. Although inter-annotator agreement is substantial, these decisions still involve interpretive judgment. Finally, we analyze disclosed vulnerabilities only and therefore do not account for undiscovered or unreported weaknesses.

\section{Conclusion and Future Work}
\label{sec:conclusion}
We empirically analyzed 295 GitHub Security Advisories referencing LLM-related components and manually annotated 100 using the OWASP Top 10 for LLM Applications 2025. Our findings show that LLM-associated packages inherit conventional weakness classes rather than introducing new ones, while existing advisory metadata lacks structured indicators of model-mediated exposure. Taken together, CWE and OWASP perspectives are complementary: neither alone is sufficient to characterize the full risk profile of LLM-integrated systems.

In future work, we will extend the dataset beyond the current one-year window and incorporate additional disclosure sources, including CVE, NVD, and ecosystem-specific advisory databases. We will also develop automated methods for identifying LLM involvement and model-mediated exposure in advisory text using the annotated dataset as ground truth. In addition, we will examine multi-stage architectural interaction pathways more systematically, especially cases where model-generated outputs propagate to downstream components that execute commands or access resources. Finally, we will conduct longitudinal and comparative analyses of how LLM-associated vulnerabilities differ from non-LLM cases in disclosure patterns, remediation timelines, and ecosystem distribution.

\balance
\bibliographystyle{ACM-Reference-Format}  
\bibliography{references}

\end{document}